\documentstyle[prl,aps,epsf]{revtex}
\newcommand{\be}{\begin{equation}}
\newcommand{\ee}{\end{equation}}

\begin{document}
\title{
Gravitating dyons and the Lue-Weinberg bifurcation}
\author{{\large Y. Brihaye, F. Grard \ and \ Sophie Hoorelbeke} }
\address{\small 
Facult\'e des Sciences, Universit\'e de Mons-Hainaut,
B-7000 Mons, Belgium }
\address{ }

\date{\today}

\maketitle
\begin{abstract}

Gravitating t'Hooft-Polyakov magnetic monopoles
can be constructed when coupling the Georgi-Glashow model
to gravitation.
For a given value of the Higgs boson mass,
these gravitating solitons exist up to a critical value of
the ratio of the vector meson mass to the Planck mass.
The critical solution is characterized by a
degenerate horizon of the metric.
As pointed out recently by Lue and Weinberg,
two types of critical solutions can occur,
depending on the value of the Higgs boson mass.
Here we investigate this transition for dyons and show
that the Lue and Weinberg phenomenon is favorized by
the presence of the electric-charge degree of freedom.

\end{abstract}
%\vfill
%\noindent {Preprint hep-th/9912023} \hfill\break
%\vfill\eject
%\newpage

\section{Introduction}

The Georgi-Glashow (GG) model consists of an SU(2) Yang-Mills field 
coupled to a (real) Higgs triplet. The self interation
of the Higgs field by means of a Higgs potential 
leads to a spontaneous breakdown of the symmetry.
One distinguished feature of this model is that
it admits topological solitons~: the celebrated t'Hooft-Polyakov monopoles
\cite{thooft,polyakov}.

By coupling the GG field theory to gravitation, one obtains the
SU(2) Einstein-Georgi-Glashow model. Several years ago, it was shown that
gravitating magnetic monopoles, as well as non-abelian black holes, exist 
in a certain domain of the space of the physical parameters 
of the model \cite{lnw,bfm1,bfm2}. For instance
for a fixed value of the Higgs boson mass,
the gravitating monopoles 
exist up to a critical value $\alpha_{\rm cr}$ of the parameter 
$\alpha$ (the ratio of the vector meson mass 
to the Planck mass). 
%%and non-abelian black holes
%%exist up to a critical value $\alpha_{\rm cr}(x_h)$ (for small enough
%%values of the horizon radius $x_h$ \cite{bfm2}).
At the critical value $\alpha_{\rm cr}$, a critical solution with a 
degenerate horizon is reached.
In particular, for small values of the Higgs boson mass,
the critical solution where a horizon first appears
corresponds to an extremal Reissner-Nordstr\o m (RN) solution
outside the horizon while it is non-singular inside.

Recently Lue and Weinberg   reconsidered the equations of the
self-gravitating magnetic monopoles
and discovered an insofar not suspected phenomenon \cite{lw}.
Indeed  for large enough values of the Higgs boson mass,
the critical solution is an extremal black hole with non-abelian hair
and a mass less than the extremal RN value.
An independent numerical analysis \cite{bhk2} of the equations has 
confirmed the results of Lue and Weinberg and reinforced the agreement
between the numerical solutions and the algebraic conditions
these solutions have to obey.

%%Exploring the transition between the two regimes,
%%occurring at some intermediate value of the Higgs boson mass,
%%Lue and Weinberg were left with a discrepancy between
%%their analytical and numerical results \cite{lw}.

During the last months,
there was a growing interest devoted to monopole and dyons,
considered as classical solutions of the non-abelian
Einstein-Born-Infeld-Higgs (EBIH) model \cite{tri1,tri2,grandi}.
This is motivated by the fact that
the low energy effective action D-brane 
is related to a Born-Infeld model (see e.g. \cite{tse}).
Because Born-Infeld models naturally contain a theta-term
 proportional to
$(F_{\mu \nu} \tilde F^{\mu \nu})^2$,
their nonlinearities are better tested by
the dyons solutions which excite both magnetic and electric fields.

It is therefore natural to study if the gravitating dyons 
of the GG model are also sensitive
to the Lue and Weinberg effect
and to see if the degree of freedom
confering the classical
lump its electric charge favorizes or attenuates this effect.
In this  report we   extend the analysis of Ref.\cite{lw} to 
the case of spherically symmetric gravitating dyons. 
We show that the transition  also occurs for dyons
and that the phenomenon is enhanced by the presence of 
the new field describing the electric charge degree of freedom.

%%%%%% ICI

\section{The equations}

We consider the SU(2) Einstein-Georgi-Glashow action
\cite{lnw,bfm1,bfm2,lw}. In order
to obtain static spherically symmetric globally regular solutions
we employ Schwarzschild like coordinates for the metric
\begin{equation}
\label{metric}
ds^2=
  -A^2N dt^2 + N^{-1} dr^2 + r^2 (d\theta^2 + \sin^2\theta d\phi^2)
\ , \end{equation}
and introduce, as usual, the mass function $m(r)$ by means of
\begin{equation}
N(r)=1-\frac{2m(r)}{r} 
\ . \label{n} \end{equation}
Then we use
the standard Wu-Yang ansatz for the spatial 
components of the gauge field and the  hedgehog ansatz for the
Higgs fields (see e.g. \cite{bfm1,lw})
\begin{equation}
\vec A_r=0 \ , \ \ \
\vec A_\theta =  -\vec e_\phi \frac{1- K(r)}{g} \ , \ \ \
\vec A_\phi =   \vec e_\theta \frac{1- K(r)}{g} \sin \theta
\ , \end{equation}
and 
\begin{equation}
\vec \phi = \vec e_r H(r) v
\ , \end{equation}
with the standard unit vectors $\vec e_r$, $\vec e_\theta$ and $\vec e_\phi$ while $v$ denotes the
Higgs field's vacuum expectation value.
Since we  want  to describe dyon, we also employ 
the spherically symmetric ansatz
for the time component of the gauge field 
\cite{jz,bhk,bhkt}.
\be
  \vec A_0  = v J(r) \vec e_r    \ \ \  .     
\ee  

It is also convenient to introduce the dimensionless coordinate $x$
and mass function $\mu$,
\begin{equation}
x = g v r \ , \ \ \ \mu=g v m
\ , \label{xm} \end{equation}
as well as the dimensionless coupling constants $\alpha$, $\beta$
\begin{equation}
\alpha^2 = 4 \pi G v^2 \ \ \ , \ \ \ \beta = \frac{M_H}{M_W} \ ,  
\end{equation}
where $G$ is Newton's constant, 
%%%$v$ is the Higgs field expectation value,
 $M_H$ is the Higgs boson mass and $M_W$ is the gauge boson mass.

With these ansatz and definitions, the classical equations of the
model reduce to the following system of five coupled radial equations~:
\cite{bhk,bhkt} 
\begin{eqnarray}
\mu'&=&\alpha^2 \Biggl( \frac{x^2 J'^2}{2 A^2} 
   + \frac{J^2 K^2}{A^2 N}
   + N K'^2 + \frac{1}{2} N x^2 H'^2
\nonumber\\
 & &\phantom{ \alpha^2 \Biggl( }
   + \frac{(K^2-1)^2}{2 x^2} + H^2 K^2
   + \frac{\beta^2}{8} x^2 (H^2-1)^2 \Biggr)
\ , \label{eqmu} \end{eqnarray}
and
\begin{eqnarray}
 A'&=&\alpha^2 x \Biggl(
    \frac{2 J^2 K^2}{ A^2 N^2 x^2}
   + \frac{2 K'^2}{x^2} + H'^2 \Biggr) A
\ , \label{eqa} \end{eqnarray}
for the functions parametrizing the geometry
(the prime indicates the derivative with respect to $x$);
for the matter functions we obtain the equations
\begin{eqnarray}
(A N K')' = A K \left( \frac{K^2-1}{x^2} + H^2 - \frac{J^2}{A^2 N} 
 \right)
\ , \label{eqk} \end{eqnarray}
\begin{equation}
\left( \frac{x^2 J'}{A} \right)' = \frac{2 J K^2}{AN}
\ , \label{eqj} \end{equation}
and
\begin{eqnarray}
( x^2 A N H')' = A H \left( 2 K^2 + \frac{\beta^2}{2}
 x^2 (H^2-1) \right)
\ . \label{eqh} \end{eqnarray}
The equations of motion depend on two physical parameters 
$\alpha, \beta$. 
This form of the parameters are used in \cite{bfm2} but 
several alternative notations were used in
previous papers. In \cite{lw} they use $a,b$ with 
\be
     a = 2 \alpha^2 \ \ \ , \ \ \ 
     b = \frac{1}{4} \beta^2 = \frac{1}{2} \tilde\beta^2 \ ,
\ee
and in passing we redefined by $\tilde \beta$ the parameter
labelled $\beta$ in \cite{bhk,bhkt}.

In solving the equations, $\alpha, \beta$ are imposed by hand;
the third physical parameter to be fixed by hand is
the electric charge of the dyon, noted $Q$. 
It is encoded \cite{jz,bkt} into the
asymptotic behaviour of the function $J(x)$
\be
\label{jas}
         J(x)|_{x \rightarrow \infty} = J_{\infty} - 
         \frac{Q}{x} + o(\frac{1}{x^2})
\ee
So that $Q$ is imposed as a boundary condition 
of the function $x^2 J'$  at $x = \infty$.

The regularity of the solution at the origin, the finiteness
of the mass and the requirement that the metric (\ref{metric})
 approaches the
Minkowski metric for $x\rightarrow \infty$
lead to the following set of boundary conditions
\cite{bhkt}
\be 
\mu(0) = 0 \ \ , \ \ K(0) = 1 \ \ , \ \ H(0) = 0 \ \ , \ \ J(0) = 0
\ee
\be
A(\infty) = 1 \ \ , \ \ K(\infty)=0 \ \ , \ \ H(\infty) = 1 \ \ ,
\ \ (x^2 J')|_{x \rightarrow \infty} = Q  \ .
\ee
 which fully specify the problem.

In absence of gravity (i.e. $a=0$, $N=A=1$)
the first two equations are trivial and  
the solutions are the dyons of Julia and Zee \cite{jz}.
They were studied numerically in some details recently
\cite{bkt}. In particular, it was found that for $b=0$ 
dyons exist for an arbitrary value of $Q$ while for
$b \neq 0$ the dyons exist only to a maximal value $Q_{cr}$
(see Fig.~8 of \cite{bkt});
for too high values of $Q$ the value 
$J_{\infty}$ becomes larger than one and,
as seen from Eq.(\ref{eqk}), $K$ becomes oscillating.
Within the range of $\beta$ that we will explore in this paper
(gravitating) dyons exist  for $Q \leq 0.7$.

The equations (\ref{eqmu})-(\ref{eqh})
 were studied in details in \cite{bhk,bhkt}
for the case $\beta=0$. In particular,
it was shown that the non-Abelian gravitating dyon
bifurcates at $\alpha =\alpha_{cr}$ into an extremal RN 
solution with
\be
      N(x) = \frac{(x - \alpha\sqrt{1+Q^2})^2}{x^2} \ \ , \ \ A(x) = 1
\label{rn1}
\ee      
\be
      K(x) = 0 \ \ , \ \ H(x) = 1 \ \ , \ \ J(x) = 
                 Q(\frac{1}{\alpha \sqrt{1+Q^2}} - \frac{1}{x})         
\label{rn2}
\ee

defined on $x \in [\alpha \sqrt{1+Q^2}, \infty]$.
In the next section we study the dyons solution for $b > 0$.

\section{Gravitating dyons}

In order to make the paper self consistent, we first
briefly recall how the magnetic monopole solutions
approach critical solutions,
when the vector boson mass, i.e.~$a$, is varied,
while the ratio of the Higgs boson mass
to the vector boson mass, i.e.~$b$, is kept fixed.
Recently, Lue and Weinberg \cite{lw} realized that there are two
regimes of $b$, each with its own type of critical solution.

In the first regime $b$ is small ($b < 25$),
and the metric function $N(r)$ 
of the monopole solutions possesses a single minimum.
As the critical solution is approached, i.e.~as $a \rightarrow a_{\rm cr}$,
the minimum of the function $N(r)$
decreases until it reaches zero at $r = r_0$.
The limiting solution
corresponds to an extremal RN black hole solution with horizon radius
$r_h = r_0$ and unit magnetic charge
for $r \ge r_0$.
Consequently, also the mass of the limiting solution
coincides with the mass of this extremal RN black hole.
However, the limiting solution is not singular
in the interior region, $r < r_0$.
We refer to this limiting approach as RN-type behaviour.

In the second regime $b$ is large,
and the metric function $N(r)$ 
of the monopole solutions develops a second minimum
as the critical solution is approached.
This second minimum arises interior to the location of the first minimum,
and decreases faster than the first minimum.
Therefore, the critical solution is reached at $r=r_*$,
where the second minimum reaches zero.
The critical solution thus possesses an extremal horizon
at $r_* < r_0$,
and corresponds to an extremal black hole with non-abelian hair
and a mass less than the extremal RN value.
Consequently, we refer to this second 
limiting approach as NA-type (non-abelian-type) behaviour.

To summarize~: the non-Abelian gravitating monopole exist on a portion
of the $(a,b)$ plane limited by a curve $a_{cr}(b)$. Somewhere on 
this curve (at $(a_{cr}=3/2,b_{cr} \sim 26.7)$) there is a
critical point separating the RN-type and NA-type of ending of the
solution.

By solving Eqs.(\ref{eqmu}) to (\ref{eqh}) we were able to
show that the same phenomenon occurs for dyon solutions. Before
to present the details of the two regimes, we discuss the solutions
in the case $Q= 1/2$, $b=50$ which is generic of the NA-type
(the corresponding critical $\alpha$ is $\alpha_{cr} \approx 0.83175$).

The profiles of the functions $A,K,H$ at the approach of the
critical value $\alpha_{cr}$ was described at length in \cite{lw}
and obey a similar pattern in the case of dyon. Therefore 
we present here the profiles of the functions $N$ and $J$. 
We found it instructive to superpose on our figures the profiles
of the functions
\be
            \frac{J}{A} \ \ \ , \ \ \ B \equiv \frac{J}{A \sqrt{N}}.
\ee
The ratio $J/A$ naturally appears when one eliminates the 
function $A$ (by using (\ref{eqa})) from the other equations.
The special combination $B$ becomes a constant when the solution
approaches an extremal RN black holes (\ref{rn1}),(\ref{rn2}).
Moreover it takes the value 
\be
\label{brn}
            B_{RN} = \frac{Q}{\alpha \sqrt{1+Q^2}}
\ee
this can be checked from the numerical solutions.

So on Figs.1-4, the functions $N, J, J/A , B$ are represented
as functions of the variable $y$
\be
          y \equiv  \frac {x}{\alpha \sqrt{1+Q^2}}   \ .
\ee
It is such that $y_0 = 1$ (with an obvious notation
we also define $y_*$ as the value of the interior minimum of N). 
Figure 1
is drawn for $\alpha = 0.75$, i.e. much before $\alpha_{cr}$.
The function $N$ still present a unique minimum around $y=1$.
Figs. 2, 3 are obtained for $\alpha = 0.8315$. Now $N$ has
developped a second minimum at $y \approx 0.80$. We observe that
$J$ (and then also $J/A$ and $B$) deviates only little from zero
for $y \in [0, 0.8]$. A little before $y=0.80$ the function $B$
increases very quickly to attain the value
(\ref{brn}) around $y=1$. We also notice that $J/A$ develops
a ripple between the two minima on $N$.
This general tendency of the functions is confirmed by Fig. 4
drawn very close to the critical $\alpha$. The function $B$
here increases very steeply (in fact quasi linearly) from zero
at $y_* \approx 0.795$ to the value (\ref{brn}) at $y=1$.

By solving the equations for several values
of $a$,$b$ and $Q$ we obtain 
a strong numerical evidence that the two behaviours found in \cite{lw}
persist in the case of dyons. This statement is illustrated by
Figs. 5,6 where the ratio 
$y_* = r_*/r_0$ characterizing the NA-type
of behaviour is plotted respectively as a function 
$Q$ (for a few fixed values of $b$)
and as a function of of $b$ 
(for $Q=0$ and $Q= 1/2$) \cite{foot}. 
The two figures indicates in particular that, when $Q$ increases,
the NA-type appears for lower values of $b$.

Corresponding to Fig. 6, the critical value $a_{cr}$ is also plotted
in Fig. 7. 
The remarkable fact indicated by this figure is that the line
$a_{cr}(b)$ in the $(a,b)$ plane is independent of $Q$
(this was also observed in \cite{bhkt} for $b=0$). 
Only the transition point $(a_{cr} , b_{cr})$
is moving with $Q$ on the line.

The fact that the curve corresponding to $Q=1/2$ overtakes
the value $a_{cr}=3/2$ is not in contradiction with \cite{lw} (i.e.
for $Q=0$). Expanding the dyon equations around a double zero
of the function $N(r)$ (i.e. $r= r_*$ or $r=r_0$) is the same way as in
\cite{lw} would lead in principle to a $Q$-depending critical value
of $a$ limiting the RN and NA-types.
Technically, it would involve the diagonalisation of a 3*3 matrix
(rather than a 2*2) and leads to a more involved algebra.
However, owing that $Q$ is related to the asymptotic decay of the
function $J$ it is very unlikely 
that an expansion of the solution around $r_*$ (or $r_0$) could
lead to an explicit expression for $a_{cr}(Q)$ and therefore we
refrain from performing such an expansion.

Clearly , Figs.6,7 report data which is obtained from a 
numerical analysis of the equation and probably the NA-type solution
(with $Q=0$ or $Q\neq0$) exists for still slightly lower values of $b$.
These solutions cannot be constructed because of the limitation
in accuracy of our numerical routine. In order to support this 
statement we report  in Fig. 8  the value $N_{min}$
of the (RN-type) minimum of $N$
 {\it when the internal minimum first appears}, 
this  as a function of $b$.
As argued in \cite{bhk2} we believe that the transition between RN and
NA-types of behaviour occurs just when this curve crosses zero.
An extrapolation of the curves gives $b_{tr} \approx 26.7$
for $Q=0$ and $b_{tr} \approx 18.7 $ for $Q=1/2$.

\section{Conclusion}

The Einstein-Georgi-Glashow model constitutes a good theoretical laboratory
for testing the properties of gravitational solitons. 
The solutions are particularly rich and obey non trivial patterns of bifurcations
of several types \cite{bfm2,bhk,bhkt}.  Here
we have analyzed numerically the equations for the spherically symmetric
gravitating dyons in this model,
focusing our attention on the intermediate region of the Higgs boson mass
parameter. 

Our results gives strong evidence of two different scenarii of ending of the 
non-abelian dyons at some critical value $a = a_{cr}(b)$, confirming the recent result of \cite{lw}. 
Three points desserve to be further stressed. 
(i) The domain of existence of gravitating dyons in the (a,b)
plane seems to be independant of $Q$. 
(ii) The transition bewtween the 
RN and NA regime occurs at a Q-depending value $(a_{cr}, b_{cr})$ of the parameters.
For monopole ($Q=0$) we have  $a_{cr} = 3/2$ , $b_{cr} \approx 26.7$.
Dyons have $a_{cr} > 3/2$ and the precise
value depends (rather indirectly) on the 
charge $Q$. (iii) At non-zero values of $b$ the dyons solutions exist up to a maximal
value $Q_{cr}(b)$, limiting the domain of investigation of the equations.

\vskip 0.5 cm
\noindent {\bf Acknowledgments}

One of us (Y.B.) gratefully acknowledge Prof. J. Kunz and 
B. Hartmann for interesting him in the topic and for their
enthusiastic collaboration in Refs.\cite{bhk,bhkt,bhk2}.

\vfill
\eject

{\bf Figure captions}

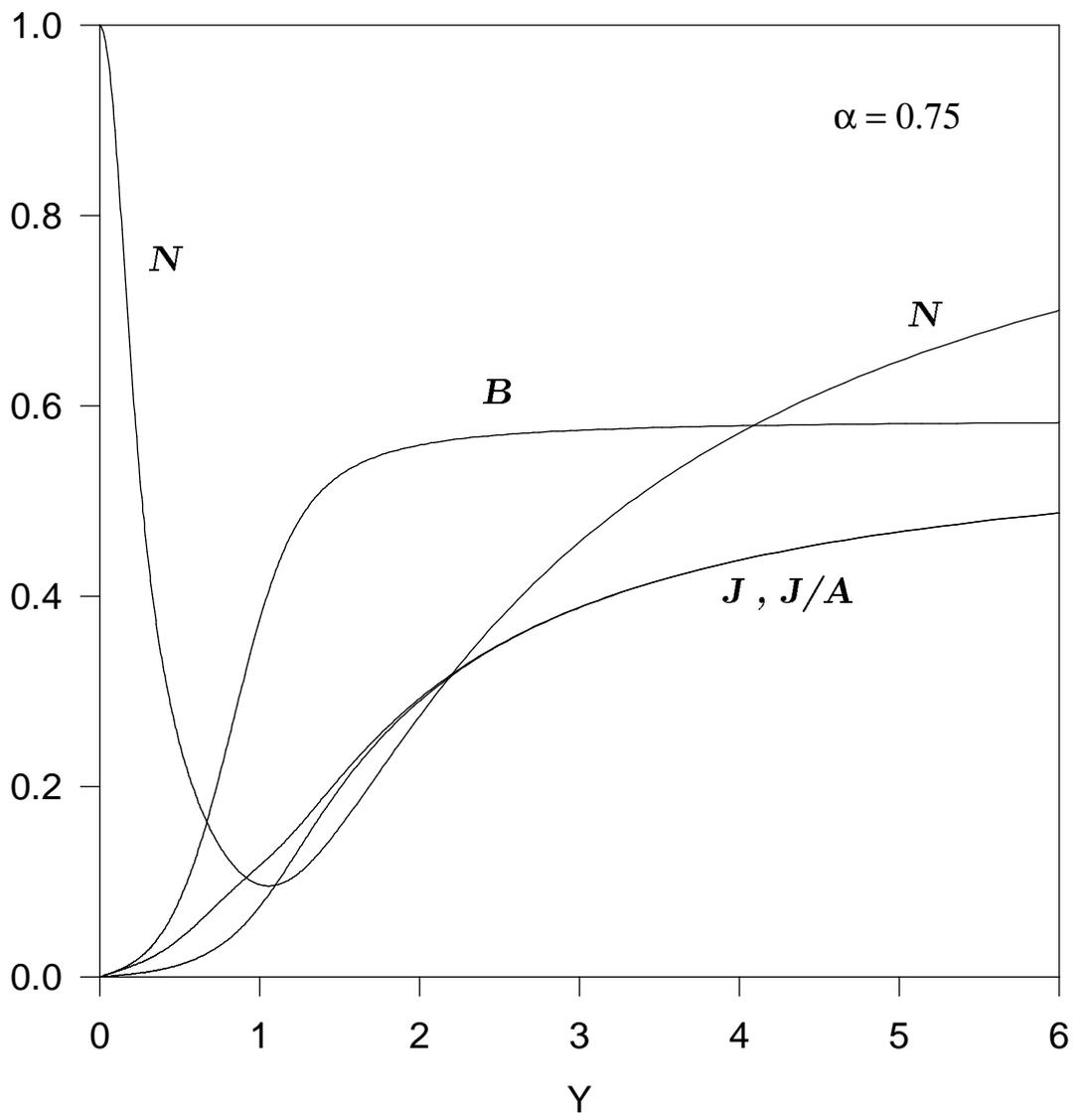
\noindent Fig.~1.

The profiles of the functions $N, J, J/A, B$ 
as functions of $y$ for $Q=1/2$, $b=50$ and $\alpha = 0.75$.

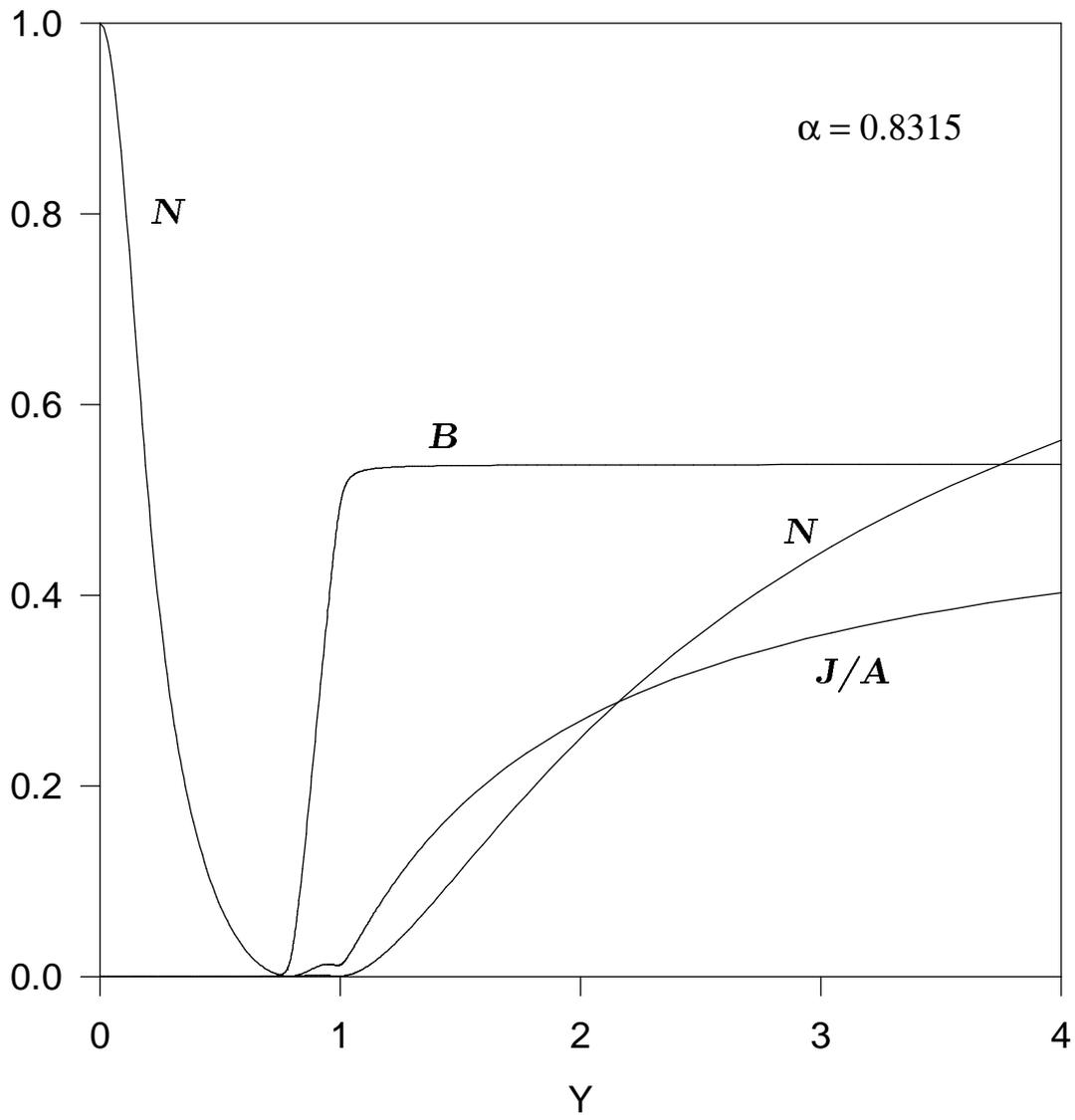
\noindent Fig.~2.

The profiles of the functions $N, J, J/A, B$ 
as functions of $y$ for $Q=1/2$, $b=50$ and $\alpha = 0.8315$. 

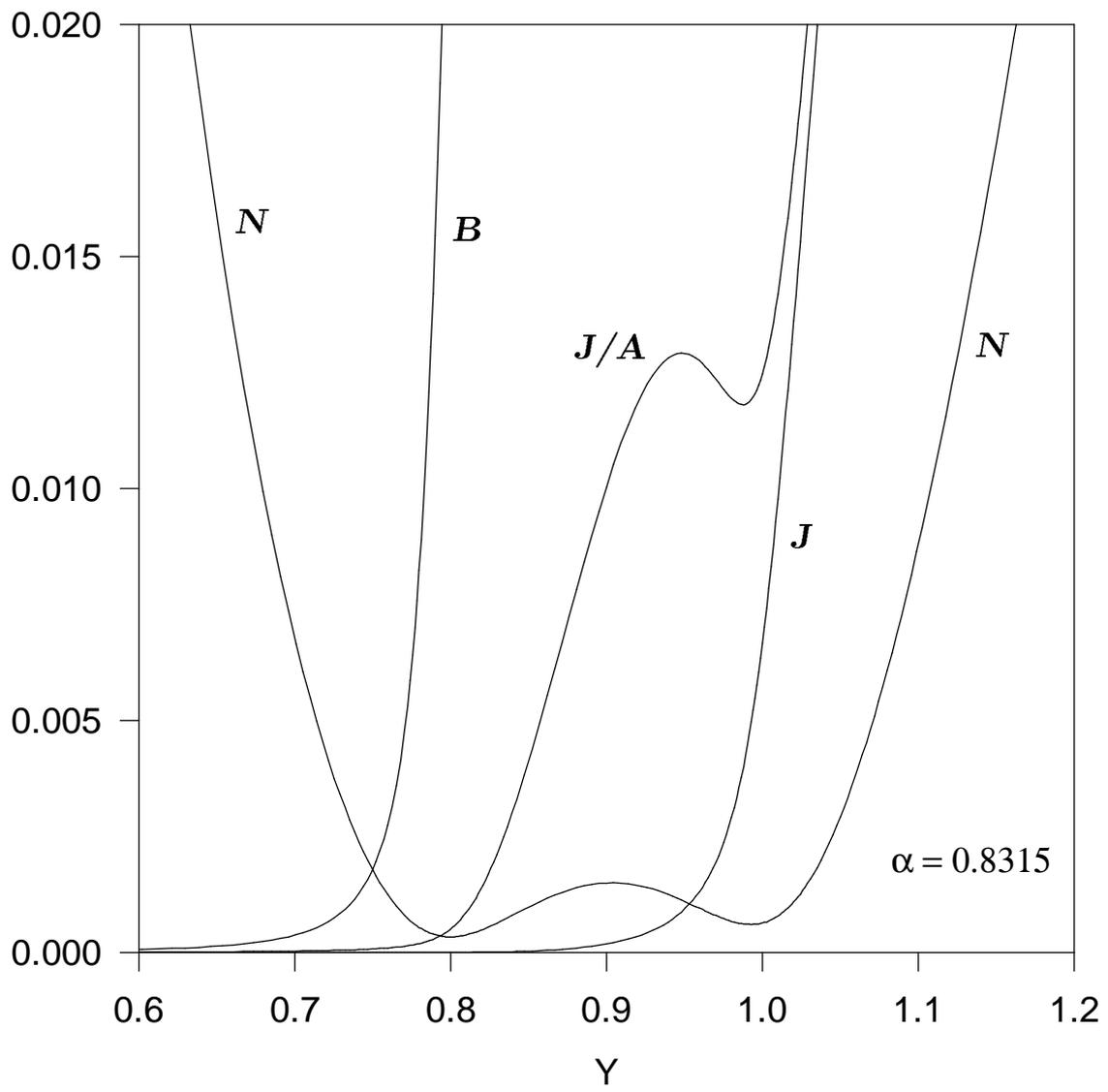
\noindent Fig.~3.

Enlarged view of Fig.~2 in the region $Y=0.9$ where the two
minima of $N$ occur. 
 
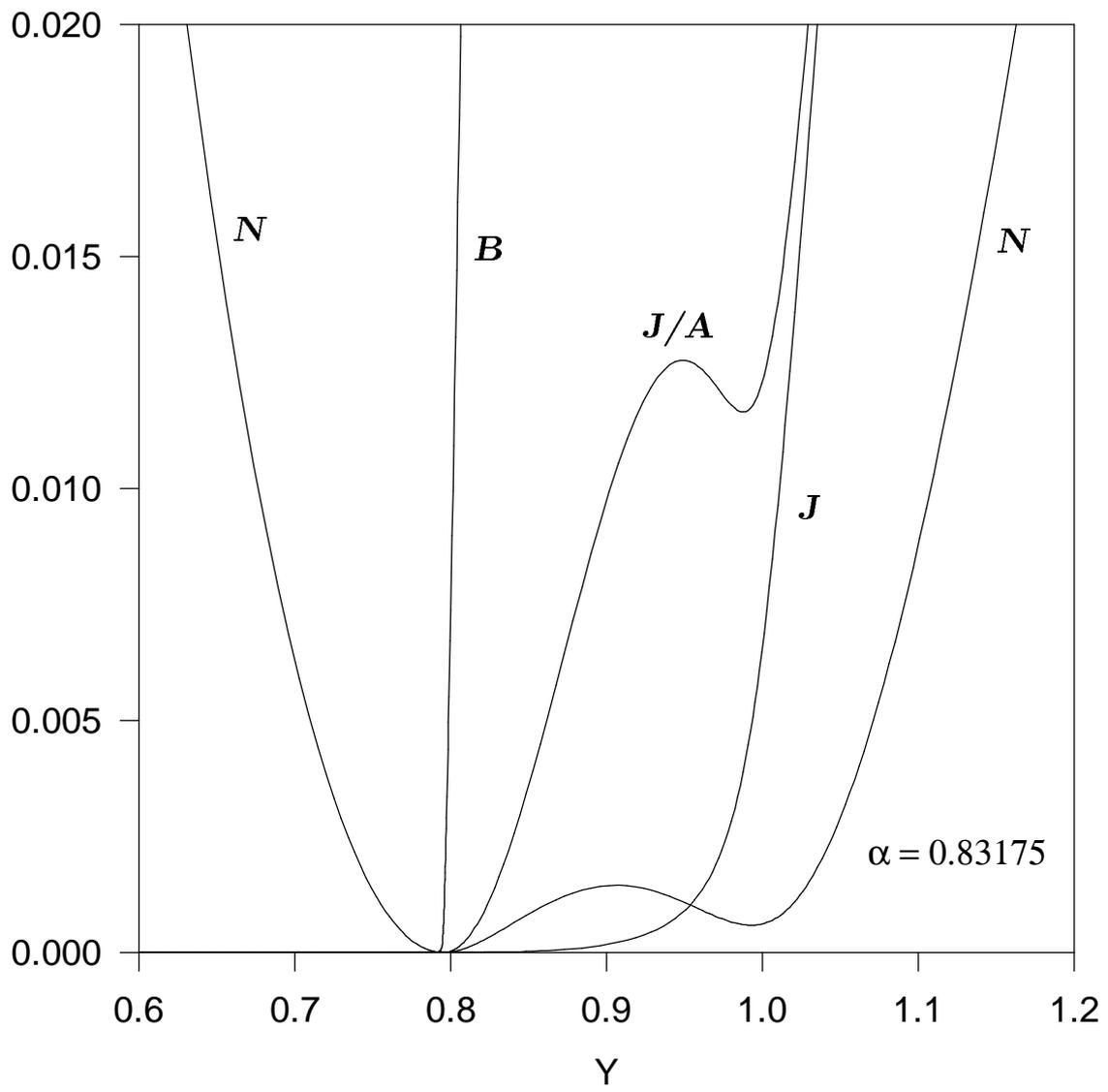
\noindent Fig.~4.

The profiles of the functions $N, J, J/A, B$ 
as functions of $y$ for $Q=1/2$, $b=50$ and $\alpha = 0.83175$. 

\noindent Fig.~5

The ratio $y_*=r_*/r_0$ is presented as a function of $Q$
in the NA-type regime for $\tilde \beta = 8$ ($b=32$)
and for $\tilde \beta = 6.5$ ($b=21.125$).

\noindent Fig.~6

The ratio $y_*=r_*/r_0$ is presented as a function of $b$
in the NA-type regime
for monopole solutions and for dyon solutions
with charge  $Q = 1/2$.

\noindent Fig.~7

The critical value $a_{\rm cr}-1$ is presented as a function of $b$
in the NA-type regime
for monopole solutions and for dyon solutions
with charge  $Q = 1/2$.

\noindent Fig.~8

The value of the function $N(r)$ at the first minimum 
where the second minimum first appears
is presented as a function of the value of $b$,
for monopole solutions and for dyon solutions
with charge  $Q = 1/2$.


\begin{thebibliography}{000}
\bibitem{foot}
The data of Figs. 6,7,8 for $Q=0$ is taken from \cite{bhk2}.
\end{thebibliography}
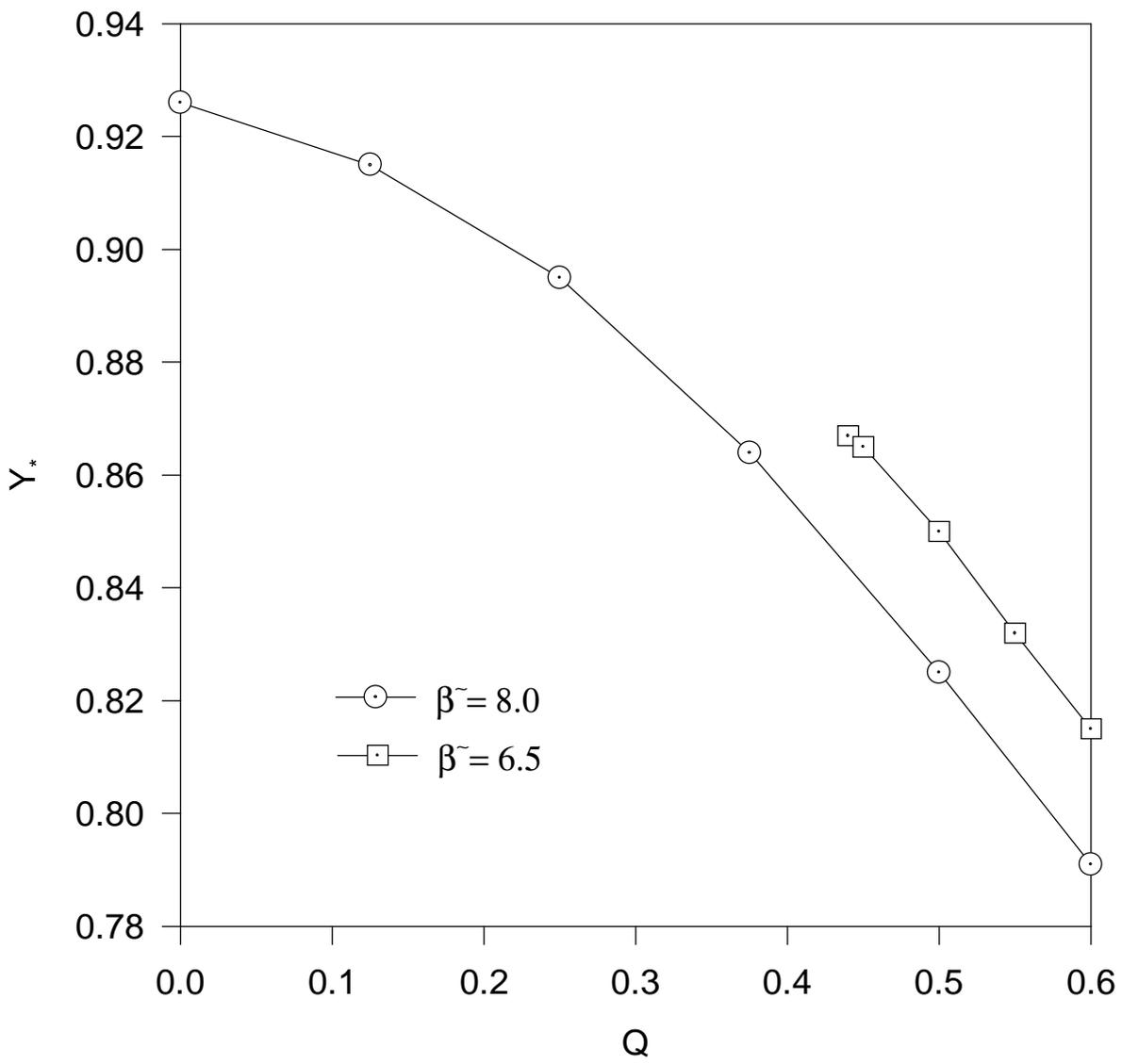
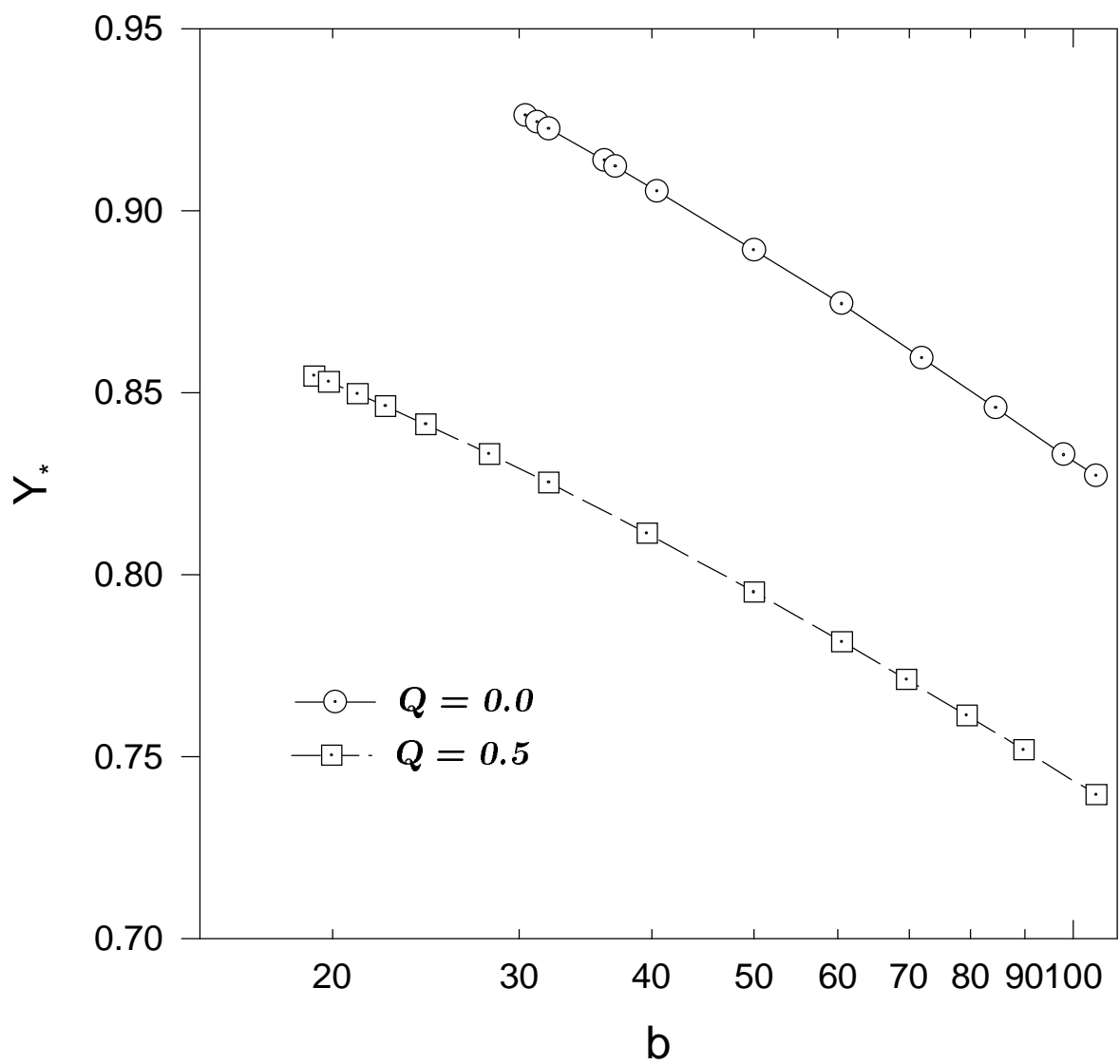
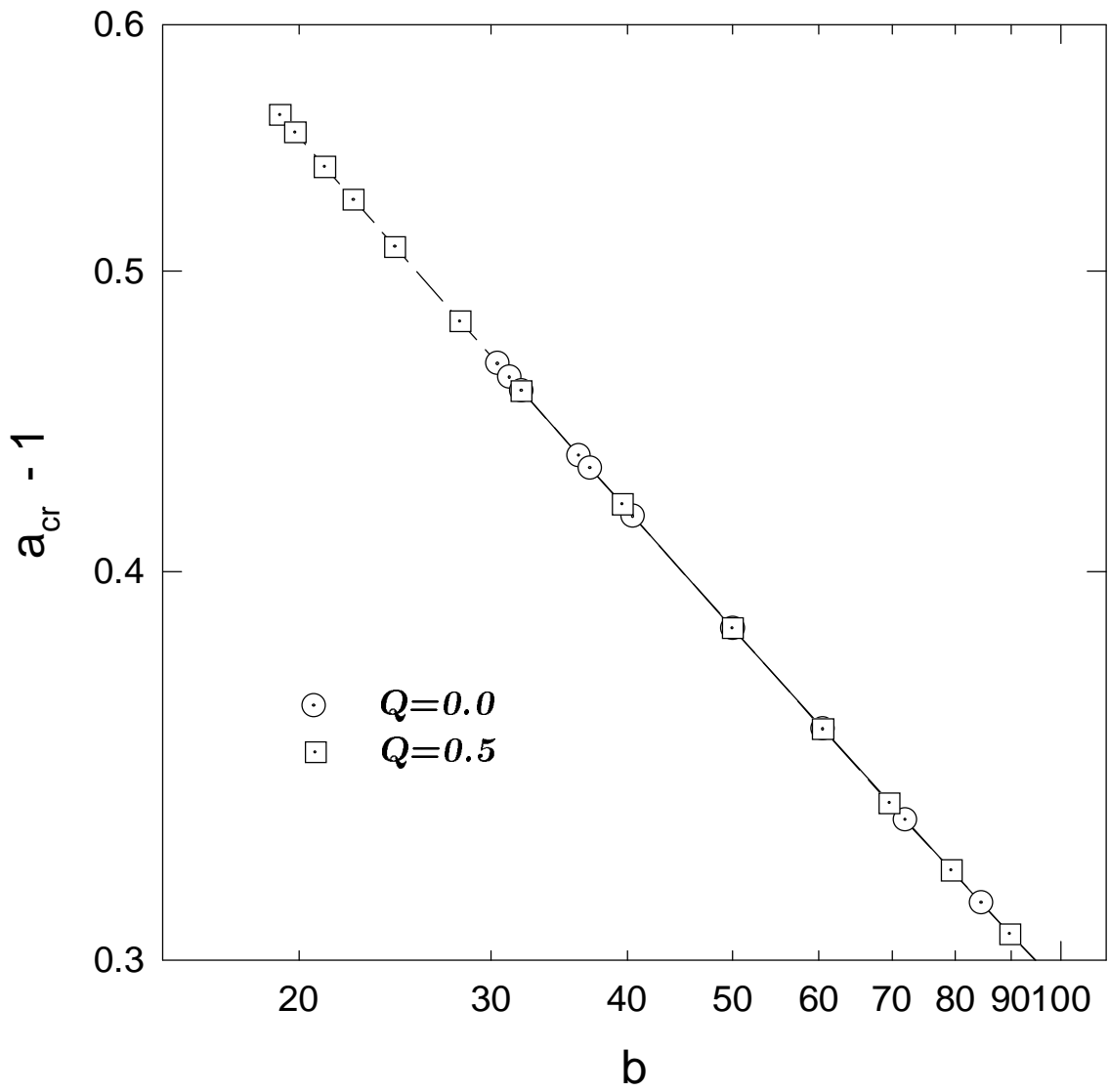
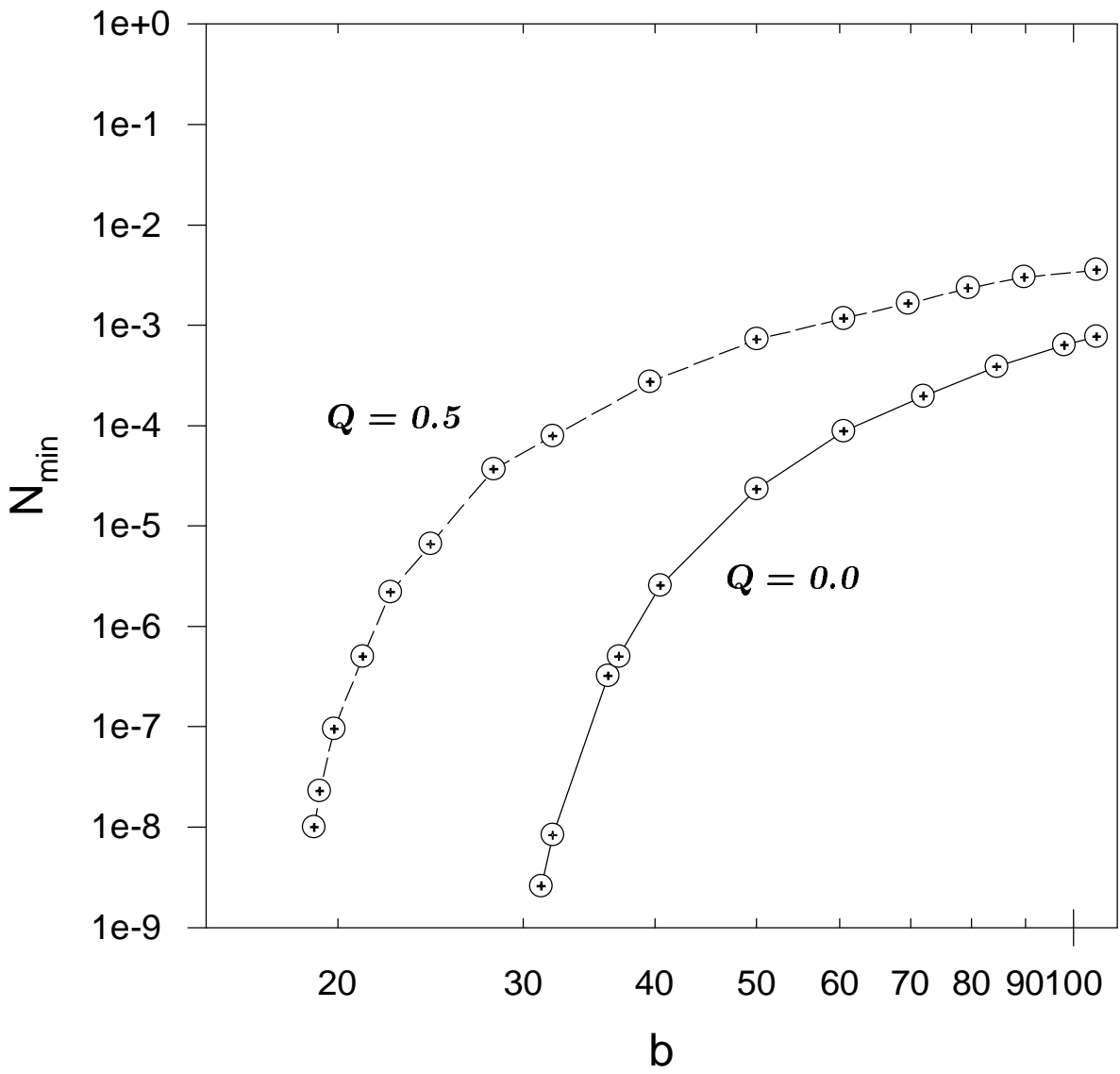
\end{document}